\documentclass[iop, tighten]{emulateapj}

\slugcomment{{\sc Accepted for Publication in AJ} }
\usepackage{amssymb,amsmath,graphicx}

\newcommand{\Mearth}{\mbox{$\rm M_{\earth}$}}
\newcommand\ie{{\it i.e.~}}
\newcommand{\twoSaturns}{{\it 2Saturns}}
\newcommand{\iceSaturn}{{\it iceSaturn}}
\newcommand{\control}{{\it control}}
\newcommand{\allIce}{{\it allIce}}

\def\about  {\hbox{$\sim$}}
\begin{document}

\shortauthors{Lewis et al.}

\title{The Influence of Outer Solar System Architecture on the Structure and Evolution of the Oort Cloud}

\author{Alexia R.~Lewis\altaffilmark{1}, 
Thomas Quinn\altaffilmark{1}, \& 
Nathan A. Kaib\altaffilmark{2}
}

\altaffiltext{1}{Astronomy Department, University of Washington, Box 351580, Seattle, WA 98195, USA; arlewis@astro.washington.edu}
\altaffiltext{2}{Department of Physics and Astronomy \& Center for Interdisciplinary Exploration and Research in Astrophysics (CIERA), Northwestern University, Evanston, IL 60208-2900}

\begin{abstract}
 We study the influence of outer Solar System architecture on the structural evolution of the Oort Cloud (OC) and the flux of Earth-crossing comets. In particular, we seek to quantify the role of the giant planets as ``planetary protectors''. To do so, we have run simulations in each of four different planetary mass configurations to understand the significance of each of the giant planets. Because the outer planets modify the structure of the OC throughout its formation, we integrate each simulation over the full age of the Solar System. Over this time, we follow the evolution of cometary orbits from their starting point in the protoplanetary disk to their injection into the OC to their possible re-entry into the inner planetary region. We find that the overall structure of the OC, including the location of boundaries and the relative number of comets in the inner and outer parts, does not change significantly between configurations; however, as planetary mass decreases, the trapping efficiency (TE) of comets into the OC and the flux of comets into the observable region increases. We determine that those comets that evolve onto Earth-crossing orbits come primarily from the inner OC but show no preference for initial protoplanetary disk location. We also find that systems that have at least a Saturn-mass object are effective at deflecting possible Earth-crossing comets but the difference in flux between systems with and without such a planet is less than an order of magnitude. We conclude by discussing the individual roles of the planets and the implications of incorporating more realistic planetary accretion and migration scenarios into simulations, particularly on existing discrepancies between low TE and the mass of the protoplanetary disk and on determining the structural boundaries of the OC. 
\end{abstract}
\keywords{Comets: general --- Oort Cloud}

\section{Introduction}
\label{sec:intro}

\cite{Oort1950} first proposed the existence of a spherical reservoir of comets at the edge of the Solar System. Because of its large distance (the semimajor axes of comets in the OC are of order 1000--200,000 AU), it is only loosely bound to the Sun and therefore extremely susceptible to external perturbations. Studying this reservoir is difficult because of this distance and the small size of the comets, and observations of long period comets (LPCs) are inherently biased because comets in the outer OC are more likely to be knocked onto observable orbits than those in the inner OC \citep{Hills1981}. Because the evolution of cometary orbits involves numerous factors, analysis of the OC, including its structure and formation, must be done with numerical simulations. Since Oort first hypothesized the OC, observations and analysis of visible LPCs and the advancement of computing capabilities have aided the development of dynamical simulations which have been utilized to dramatically enhance our understanding of the structure and formation of the OC. 

One of the current drivers behind the study of the OC is to understand the terrestrial bombardment rate and its impact on planetary habitability. This is directly influenced by the structure of the OC; the number of comets and their locations in the OC will affect the size of any comet shower, a 2--3 Myr period of elevated cometary flux due to an extremely close stellar passage \citep{Hills1981}. In order to understand this impact rate, we must quantify the role played by the outer planets in shielding the inner Solar System from this flux of particles, including the evolution of cometary orbits as a result of perturbations by the outer planets. Existing studies have looked at parts of this question. \cite{Wetherill1994} calculated perihelion passages of short-period comets from the Kuiper belt in planetary systems similar to our own but with different mass arrangements, including one that mimics the present system, one that contains ``failed Jupiters" of 15 \Mearth, and one in which Neptune and Uranus have masses of 1 \Mearth. The number of passages calculated in the ``failed Jupiters'' case is two to three magnitudes higher than in the present system. In an ongoing series of papers \citep{Horner2008, Horner2009, Horner2010}, the authors examined the role played by Jupiter in shielding the inner Solar System from various populations of small bodies: the asteroids, the Centaurs, and the OC comets. In the first two papers, they found that systems with no Jupiter actually had fewer impacts than systems that had a mid-sized Jupiter. In the third paper of the series, \cite{Horner2010} model the population of Oort cloud comets over 100 Myrs under the effect of Jupiter, Saturn, and the Sun. For this population of small bodies, they find that Jupiter does indeed act as an effective shield, and its effectiveness increases with mass. 

These works have important limitations. \cite{Wetherill1994} was interested in determining an approximate impact rate on Earth, but his numbers are more representative of Kuiper belt objects than OC objects. He did not explicitly include Galactic tides, but rather made assumptions regarding the passage of particles from the planetary region to the OC. As a result, his calculations most accurately describe the effect on the impact rate of short-period comets and are more appropriately compared to the results of \cite{Horner2009}. \cite{Horner2010} look specifically at LPCs originating in the OC, though they follow the evolution of the particle orbits for only 100 Myrs and under the influence of Jupiter and Saturn, and do not explicitly model passing stars, the Galactic tide, and other external perturbers. 

In order to quantify the impact rate at Earth from OC objects, we must accurately understand the formation of the OC and the manner in which objects are sent to the OC, perturbed out of it, and ejected from the Solar System. This process is complex because it depends on many factors. The OC was shaped during the formation of the Solar System, primarily by the outer planets, which are responsible for either placing particles onto bound OC orbits or entirely ejecting them from the system \citep{Fernandez1978}. It was also molded by external forces such as passing stars and the Galactic tide \citep{Byl1983, Heisler1986, Brasser2008, Fouchard2011}, and the dynamics of each of these events must be included in realistic models. The solar birth environment and its dynamical evolution in the Milky Way will alter external perturbations and thus the evolution of the OC will not be uniform with time \citep{Brasser2006, Kaib2011b}.

The increasing complexity of numerical simulations has allowed for more complete analysis of these factors. Starting with \citet{DQT1987}, numerical simulations included perturbations from the giant planets, passing stars, and the Galactic tide. Previous studies have explored the effect of a changing Galactic tide \citep{Matese1995,Brasser2010}, finding that it mainly affects OC structure but not the TE of objects into the OC. \cite{Fernandez1997} examined the effects of different solar birth environments and a number of more recent studies have looked at the influence of the Sun's birth cluster \citep[e.g.,][]{Brasser2006, Levison2010, Brasser2012}. \cite{Kaib2008} modeled the formation of the OC in various solar birth cluster environments, finding that although TE does increase with cluster density, it is mostly offset by larger numbers of close stellar encounters that strip comets from the outer regions of the cloud. \cite{Brasser2007} included the primordial solar nebula and found that its presence hinders OC formation and constrains the size of objects that can become OC comets. \citet{Kaib2011b} examined the properties of the OC in the context of an evolving solar neighborhood, accounting for changes in the Galactic tide and passing stars as a result of solar migration. They found that solar analogs that reached their smallest galactocentric distance late in their orbital evolution accumulated the smallest OC populations due to greater erosion of the outer OC in the denser regions closer to the Galactic center.

These studies reveal how the OC is shaped and altered by the Solar System and by the Galactic environment in which it is located. As the system forms, comets are continually sent into and perturbed out of the OC; its population is in a constant state of change. As a result, it is clear that in order to accurately model the OC, one must include all of these perturbing forces and model how they change with time. This must be done over the entire age of the Solar System.

Although the Sun's Galactic environment is integral in shaping the OC and stimulating or suppressing LPC flux, an equally important role is played by our Solar System's giant planets. In this paper, we examine the effect of planet mass on the structure of the OC and the flux of comets into the inner Solar System. The magnitude of planetary perturbations is dependent on the mass of the perturbing planet---the more massive the planet, the greater the perturbation. Previous work that has examined the change in the impact rate as a function of planetary mass has looked at just the impact when changing the mass of Jupiter \citep{Wetherill1994, Horner2008, Horner2009, Horner2010} because as the most massive planet it is assumed to have the largest effect. No study has examined the implications of varying the mass of the other outer planets, nor has existing work illustrated the long term effect on the evolution of LPC orbits and resultant OC structure. In Section~\ref{sec:methods}, we discuss the model used to simulate each OC and our method of creating a statistically significant data set with which to study the flux of comets into the inner Solar System. We present the results of the simulations in Section~\ref{sec:result} and discuss the implications of these results, particularly as they pertain to the roles of individual planets, in Section~\ref{sec:discussion}. We summarize our conclusions in Section~\ref{sec:conclusions}.

\section{Numerical Methods}
\label{sec:methods}

To study the effect that the giant planets have on OC structure and cometary flux through the inner Solar System, we follow the orbits of comets from the initial planetary disk to the present day. We model this process using the integration package SCATR \citep[Symplectically Corrected Adaptive Timestep Routine,][]{Kaib2011a}. The comets are represented by massless, non-interacting test particles, placed on random orbits and integrated over 4.5 Gyrs under the influence of the four giant planets, passing stars, and a static Galactic tide. While the orbits of the planets have certainly evolved over the age of the Solar System, we do not account for this variable in order to isolate the effect of varying planet mass. Finally, in order to accurately model the flux of observable comets, we run a set of simulations in which the particles are cloned at certain critical points in their orbits. In this way, we construct an OC that contains enough comets such that a statistically significant number will enter the inner Solar System.

SCATR is an adaptive timestep code based on routines found in the SWIFT RMVS3 integrator package \citep{Levison1994} that makes use of a symplectic corrector to reduce numerical error and increase accuracy. An adaptive timestep is particularly useful for simulations of OC formation, in which particles are both scattered by the planets and torqued by passing stars and the Galactic tide, because it allows for efficient integration over a large range of timescales. Rather than integrating the entire simulation with the small timestep required of interactions with the planets, we can run the simulation much more efficiently by adjusting the timestep when particles are no longer interacting with the giant planets. Particles on orbits inside 300 AU are integrated in the heliocentric frame with a timestep of 200 days, allowing for the increased resolution needed when accounting for planetary perturbations on the test particles. When test particles reach distances greater than 300 AU from the Sun, they are integrated in the barycentric frame using a 9000 day timestep, significantly speeding up the run time. Additionally, RMVS3 routines are well suited to this type of work because they can accurately integrate orbits to very high eccentricity.

Passing stars are created following \cite{Rickman2008}. Thirteen classes of stars are chosen. Each class is assigned a characteristic mass and a relative encounter frequency as in \cite{Garcia2001}. The encounter velocity is calculated by finding the peculiar and heliocentric velocities and assigning each velocity vector a random relative orientation. The time of each stellar passage is randomly chosen from a uniform distribution within the range [0, 4500] Myrs, which results in approximately 10 stellar passes per Myr. Finally, we create a distribution of impact angles relative to the point of entry for each star. Stars are assigned a random direction of motion as they enter the sphere, creating a distribution over the range $[-\pi/2, \pi/2]$. We then rotate the position and velocity vectors about the z- and x-axes by a random angle, creating a three-dimensional sphere of stellar encounters. Each encounter begins at a distance of 1 pc.

The Galactic tide is modeled as in \cite{Levison2001} with a rectangular coordinate system ($\tilde{x}, \tilde{y}, \tilde{z}$) centered on the moving Sun. $\tilde{x}$ points away from the Galactic center, $\tilde{y}$ points in the direction of Galactic rotation, and $\tilde{z}$ points towards the south Galactic pole. The acceleration on a test particle due to Galactic tides is then 
\begin{equation}
  \begin{split}
    F_{tide}& = \left(A-B\right) \left(3A+B\right)\, \tilde{x} \, \hat{\tilde{x}} - \left(A-B\right)^2\, \tilde{y} \, \hat{\tilde{y}} \\
    & - \left[4 \pi G \rho_0 - 2 \left(B^2-A^2\right)\right]\, \tilde{z} \, \hat{\tilde{z}},
  \end{split}
\end{equation}
where $A$ and $B$ are the Oort constants, $\rho_0$ is the density of the Galactic disk in the solar neighborhood, and $G$ is the gravitational constant. We have adopted $\rho_0 =$ 0.09 $M_{\odot}$ pc$^{-3}$.

Each simulation is integrated over 4.5 Gyrs. In this time, most of the particles are lost. This occurs in a variety of different ways. Particles are removed from the system if they collide with the Sun or one of the planets. Collision with a planet occurs if the particle enters within the radius of the planet. Collision with the Sun occurs when the particle comes within the user-specified stopping radius. Particles may also be removed from the system if they reach a distance greater than 200,000 AU from the central body.

\subsection{Primary Simulations} 
We run four simulations with different planetary mass configurations. In the first setup, all of the planets have their current masses (\control). In the second, we reduce Jupiter's mass to that of Saturn (\twoSaturns). In the third configuration, we reduce Saturn's mass by a factor of 5 (\iceSaturn). Finally, we reduce the mass of each planet to that of Neptune (\allIce). The planets are randomly placed on their current orbits. Although changing the mass of each planet will alter its orbit as well as the orbits of the other planets, in an effort to minimize the number of variables in the problem, we use the same initial conditions in all simulations. 

Each run begins with 100,000 test particles that are initially placed on random orbits with semimajor axes ($a$) between 4 and 40 AU distributed assuming a disk surface density that goes as $r^{-3/2}$. Initial eccentricities ($e$) vary uniformly between 0 and 0.01 and the cosine of the initial inclinations ($i$) varies uniformly between 0 and 0.02.  The remaining orbital elements (argument of perihelion, $\omega$, longitude of ascending node, $\Omega$, and mean anomaly, $M$) are chosen from a random distribution in the range $[0,2\pi]$. In order to decrease computing time, we divide the full simulation into 100 groups and run them separately, combining the results at the end. In these 100 groups, we use 10 different input planet files in which the initial positions and velocities of the planets along their current orbits are randomly assigned. This is done in order to avoid being dominated by chance mean motion resonances. We have checked the orbital elements of each planet and observe that they evolve in a similar fashion for each group. Stellar encounters are identical across simulations; that is, particles encounter the same stars in the same order and at the same time. We refer to these simulations as the primary simulations.

\subsection{Cloned Simulations} 
By the end of the primary simulation, most of the particles will be ejected from the system due to close encounters with and resultant scattering by the giant planets. Only a small number will have been scattered into the semimajor axis range where they can be torqued out of the planetary region by passing stars or the Galactic tide and onto safely bound orbits. Of those particles that become OC objects, only a small number will be on orbits that bring them into the inner Solar System. In order to decrease the uncertainty in the flux of comets into the inner Solar system, we need to increase the number of particles. We run four additional simulations of 200,000 test particles. Because we are interested in the flux of OC objects (those with $q > 45$ AU and $a > 300$ AU), we want to create our initial sample from a distribution of particles that contains a large number of OC objects. The cloned simulations therefore involve a two-step process: creation of the initial distribution followed by a run of the full simulation. 

To create the initial distribution, we run a 300 particle simulation using the cloning routine in SCATR. The setup of the 300 particle simulation is the same as that for the primary simulations, just with fewer particles. When cloning is turned on, clones are generated at two points in the evolution of particle orbits. The first time a particle reaches $a > 100$ AU or $q > 45$ AU, 9 additional particles are created, each with its cartesian coordinates randomly shifted from that of the original particle by $\pm 1 \times 10^{-7}$ AU. The particles are massless, but each is assigned a mass fraction, initially equal to unity. When cloning occurs, the 10 resulting particles each have a mass fraction of 0.1. Each of these particles may be cloned again, such that the smallest allowable mass fraction that a particle may have is 0.01. In this way, each particle has the potential to result in 99 additional particles, producing up to 30,000 particles from our initial 300. We run the cloning simulation for the full 4.5 Gyrs. The test particles that remain at the end of the simulation consist primarily of OC objects. 

With this distribution of OC objects, we run a 500 Myr simulation in each planetary mass configuration, which we will refer to as the cloned simulations. The initial 200000 particles are randomly selected from those test particles that survived to the end of the cloning run. The values for $a$, $e$, $i$, $\omega$, and $\Omega$ of each particle are unaltered from the sampled particle, and $M$ is randomly generated from a uniform distribution. A single particle may be sampled more than once, but the resulting orbits of the new particles will be different because the mean anomalies are all different. The resulting particles are then integrated over 500 Myrs under the same conditions as in the primary simulations. We execute the cloned simulations in order to better understand the flux of OC objects into the inner Solar System (inside 2 AU). Because most of the test particles started out as OC objects, the flux that we measure will be that of OC objects. These simulations will be used in Section \ref{subsec:flux}.

\section{Simulation Results}
\label{sec:result}

We follow particles from their initial location in the protoplanetary disk to their injection into the OC to their present day Earth-crossing orbits. This occurs in many stages during all parts of the overall evolution of the Solar System. The particles must first interact with the planets while they are still in the protoplanetary disk and their perihelion is small. Some of these particles will be ejected from the Solar System on unbound orbits while others will be sent to the OC. While in the OC, particles are subject to disruption by the Galactic tide and passing stars. A particle's location in the OC determines how these disturbances will affect its orbit. Particles in the outer part of the cloud are very loosely bound and are most likely to be ejected while those in the inner part of the cloud may be perturbed onto orbits that send them back into the planetary region and possibly into the inner Solar System. The process then repeats itself, though each pass through the planetary region means the particle is more likely to be ejected entirely from the system. 

To further understand this process, we examine the physical mechanisms that connect the initial outer planetary architecture to the ultimate flux of OC objects through the inner Solar System. The TE of particles into the OC gives us a relation between the number of particles in the OC and the initial mass of the protoplanetary disk around the Sun. Examination of OC structure through analysis of the distributions in $a$ and $q$ of OC particles tells us about the relative importance of planetary, as opposed to external, contributions to molding the OC. 

We must first define what we mean by an ``Oort Cloud'' object. We follow \citet{Kaib2008} and define an OC object as a particle with $q > 45$ AU. At this distance, the planets no longer play a significant role in the evolution of the test particle orbits. The particles do not receive additional energy kicks from the planets, so they are in safely bound orbits. Some studies \citep{DQT1987,Dones2004,Brasser2010} have included cuts in semimajor axis to further refine the definition of an OC object, specifically defining a minimum semimajor axis that delimits the inner edge of the OC. In this paper, we include a lower limit at $a=300$ AU to avoid particles that have been pumped up due to Kozai/MMR dynamics \citep{Gomes2005_resonance}. In Section \ref{subsec:timescales}, we also examine the traditional analytical arguments used to calculate the inner boundary in an attempt to determine how the planets affect this location.

\subsection{Trapping Efficiency}
\label{subsec:trapeff}

We begin by looking at TE, the fraction of initial particles that reside in the OC at the completion of each simulation. Particles become OC objects as a result of a two-step process. They first must have their semimajor axes elongated, which results from perturbations as they pass through the planetary region. As $a$ increases, they become less tightly bound to the Sun and are consequently more susceptible to perturbations from external sources. These interactions can lift the perihelion of particles out of the planetary region so they end up on bound OC orbits rather than ejected into interstellar space.

When particles encounter the planets, they receive a transfer of energy that is proportional to $M_p/a_p$, where $M_p$ is the mass of the perturbing planet and $a_p$ is its semimajor axis \citep{Dones2004}. Particles that interact with Jupiter or Saturn are often expelled from the Solar system because they receive energy kicks that are so large that they often overshoot the energy window of the OC and instead become unbound. Rather, most particles reach the OC as a result of perturbations while passing through the Uranus-Neptune region. They receive a smaller energy kick from these planets. This smaller energy change is enough to allow diffusion in $a$ to a distance where external perturbers can torque particles to larger perihelion distances, out of the planetary region and safely into the OC. 

The larger masses and smaller $a$ for Jupiter and Saturn result in energy changes that are one to two orders of magnitude greater than those from Uranus and Neptune. While it is certainly the combination of these two parameters that results in the magnitude of the energy change, an adjustment in just the mass of the planet could have a substantial effect on particle orbits. We do this in the \twoSaturns\ and \iceSaturn\ runs to determine if particles that interact with the reduced mass planet have a greater chance of making it to the OC and remaining bound to the system because they receive a smaller kick as a result of that interaction. 

\begin{figure}[!t]
  \begin{center}
    \plotone{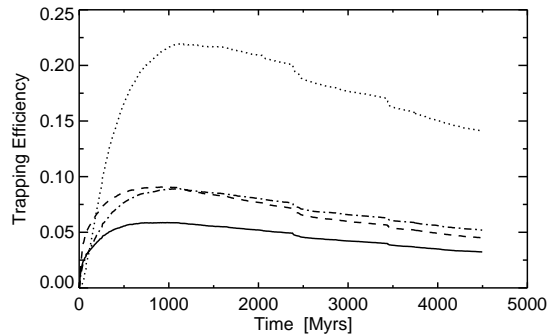}
  \end{center}
  \caption{The trapping efficiency of particles into each cloud as a function of time. At any given time, the trapping efficiency is given by the fraction of total particles that are currently defined as OC objects with $q > 45$ AU and $a > 300$ AU. Solid: \control, Dashed: \twoSaturns, Dash-dot: \iceSaturn, Dotted: \allIce}
  \label{fig:OCtrapEff}
\end{figure}

In Figure~\ref{fig:OCtrapEff}, we plot the TE as a function of time for each mass configuration in the primary simulations. The evolution of the TE is roughly the same for each OC: it steadily rises to a peak, which occurs at \about1000 Myrs in each case and then gradually decreases over the next 3.5 Gyrs to its final value. The relative efficiency between simulations provides clues as to the effect of the giant planets. Using the \control\ run as the reference, we see that the TE rises steeply over the first 100--200 Myrs before tapering off to the peak. Between its peak value and the end of the simulation, the \control\ OC loses 45\% of its particles. The \iceSaturn\ run traces the \control\ TE for the first 100 Myrs and then continues to rise as the \control\ TE begins to level off to its peak. In the \twoSaturns\ run, the TE rises much more steeply to its peak. The \allIce\ run has the most efficient population of its OC over the first 1000 Myrs though its initial population is delayed by \about30 Myrs compared to that of the other simulations. Despite this delay, by 200 Myrs it has succeeded in sending more of its particles to the OC than the other simulations. The peaks all occur at \about1000 Myrs with the \control\ and \twoSaturns\ run reaching that mark slightly before that time and the \iceSaturn\ and \allIce\ runs occurring slightly after that time. Once the peak occurs, each OC gradually begins shedding particles as they interact with passing stars and enter the planetary region. The \twoSaturns, \iceSaturn, and \allIce\ runs lose 50\%, 41\%, and 35\% of their particles, respectively, between 1 Gyr and 4.5 Gyrs. 

The \control\ run has the lowest final efficiency of 3.2\%. This is similar to recent results by \citet{Brasser2010} who find an overall efficiency of 4\% and \citet{Dones2004}, who recover \about5\%. The range in trapping efficiencies in other studies varies widely though. \citet{Brasser2006} report a range in efficiency from 2-18\%, depending on the initial density of the Solar birth cluster. Looking at just the outer OC after 1 Gyr, \citet{Dyb2008} find an efficiency of no more than 1\% and \cite{DQT1987} find that \about24\% of their initial particles are found in the OC at the end of their simulations. \citet{Dones2004} point out that this result may be inaccurate because \citet{DQT1987} initialized their particles on low-inclination, high-eccentricity orbits. This sets perihelion distances early. As a result, particles in the Uranus-Neptune region were unlikely to evolve inward and experience perturbations from Saturn or Jupiter, making them more likely to end up in the OC.

The \twoSaturns\ and \iceSaturn\ runs have similar efficiencies, both slightly higher than the \control\ at 4.5\% and 5.2\%, respectively. In these two configurations, we have reduced Jupiter to a third of its current mass and Saturn to one fifth of its current mass, individually. The two inner planets have small enough semimajor axes that changes to their masses would have to be substantial to result in a large increase in the number of OC objects. If,on the other hand, the mass changes are more substantial, the increase in TE reflects those changes. In the \allIce\ case, we set the mass of all planets equal to that of Neptune. This reduces the mass of Jupiter to \about1/20 of its current mass, a change which alone could have a significant effect on the resulting TE. When combined with a reduction in the mass of Saturn by a factor of \about5 (and a small increase in the mass of Uranus), the TE jumps considerably to 14.1\%. 

The necessity of running the simulation for the entire 4.5 Gyr history of the Solar System is clear. The TE evolves over this entire length of time so in order to understand the effect of the planets on shaping the OC, the system must be allowed to fully evolve.

\subsection{Inner-to-Outer OC Ratio}
\label{subsec:IOratio}

Once particles become OC objects, their location within the OC becomes of interest because it determines the perturbation mechanisms that affect each particle and the evolution of its orbit. While \citet{Oort1950} originally concluded that the peak in the semimajor axis distribution at 20,000 AU indicated the inner edge of the comet cloud, he did note that stellar passages that could affect comets with smaller semimajor axes were infrequent enough that observations could not say anything about the number of comets at those distances. \citet{Hills1981} later suggested the existence of an inner cloud with a mass as much as two orders of magnitude greater than that of the observed OC and that comets that enter the inner Solar System in bursts and showers came from this more massive inner cloud. The observed edge at 20,000 AU became the boundary between the new inner and outer OCs.

\begin{figure*}[!Ht]
  \begin{center}
    \plotone{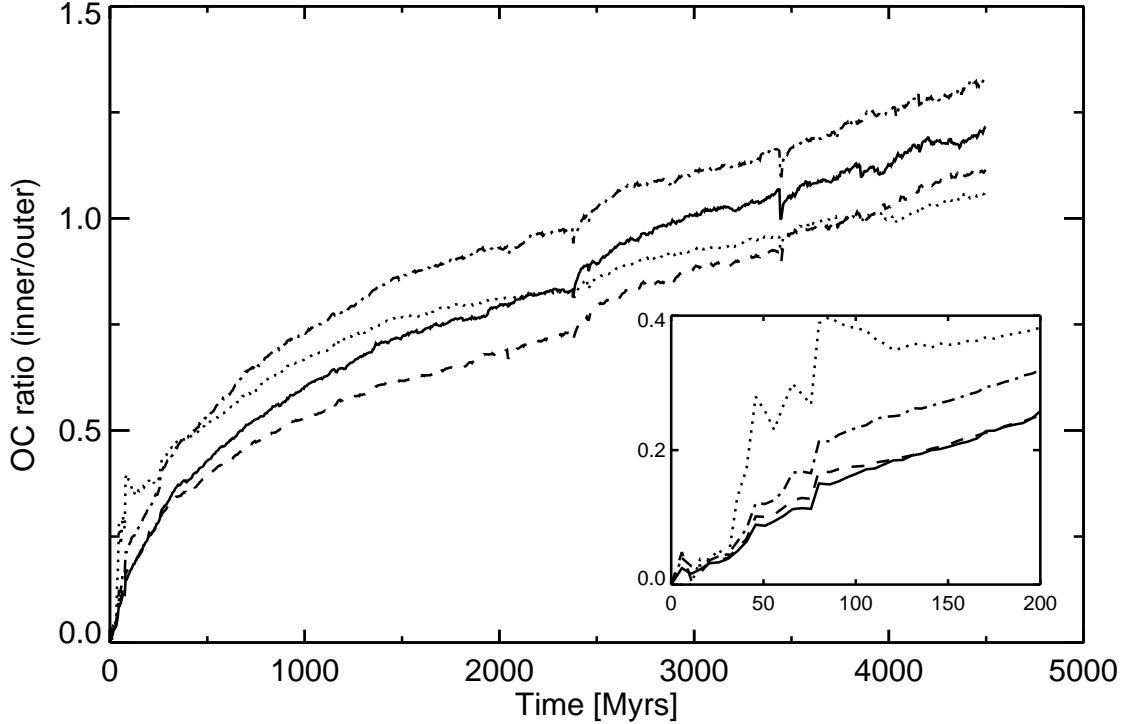} 
  \end{center}
  \caption{Ratio of the number of particles in the inner OC (300 AU $< a <$ 20,000 AU) to the number of particles in the outer OC ($a >$ 20,000 AU) as a function of time. Inset: The first 200 Myrs. Solid: \control, Dashed: \twoSaturns, Dash-dot: \iceSaturn, Dotted: \allIce}
  \label{fig:OCratio}
\end{figure*}

The relative population of the inner and outer OC is, consequently, an indicator of the threat of impacts. A more massive inner OC represents a larger threat because a passing star that generates a comet shower has the ability to perturb more comets. To assess this relative population, we split each of our resulting OCs into an inner OC and an outer OC (inner: 300 AU $< a <$ 20,000 AU, outer: $a > 20,000$ AU) and compare the number of objects in each region. Figure \ref{fig:OCratio} shows the ratio of the population of the inner OC to that of the outer OC. The end ratio in each mass configuration ranges from 1.06 in the \allIce\ configuration to 1.33 in the \iceSaturn\ configuration.

The evolution of this ratio is similar for the \control, \twoSaturns, and \iceSaturn\ configurations. As Figure~\ref{fig:OCratio} suggests, OC particles initially occupy more extended orbits in the outer part of the cloud. During this time, perturbations from passing stars disrupt these outer OC particles, usually unbinding them from the system. At the same time, many particles are still interacting with the planets, and they are being torqued onto more tightly bound OC orbits. Eventually, the populations of both parts of the comet cloud decrease, though the outer cloud begins to lose particles earlier than the inner cloud, leading to a steady increase in the ratio. In the first 100 Myrs, the \allIce\ configuration shows two sharp increases, which are more clearly illustrated in the inset of Figure~\ref{fig:OCratio}. These increases, as well as the sudden decreases that occur later in the evolution, are indicative of passing stars, and they occur at the same point in all simulations because each run uses the same star file. The lack of a massive planet in the \allIce\ run makes the system more reactive to passing stars and the peaks are larger. The dip at $t=3445$ Myrs appears very strongly in the \control\ run because the decrease in the number of inner OC objects is magnified by a brief increase in the number of outer OC objects. This behavior is seen in the \iceSaturn\ run as well, which also shows a more dramatic decrease in the ratio between the two populations. The relative dip is smaller in the \allIce\ case because of its increased number of particles, so a passing star can disrupt a larger number of particles without significantly altering the structure.

Although the general trend is the same among the four simulations and the final ratio between the inner and outer populations of each OC are comparable, there are distinct differences between the ratios of each OC that not only affect the structure of the OC as it evolves, but may also affect the number of particles that are perturbed back onto planet-crossing (and potentially Earth-crossing) orbits.

\subsection{Oort Cloud Structure}
\label{subsec:OCstruct}

In this section, we investigate the morphology of each OC to determine if different planetary mass combinations result in OCs that are structurally distinct. Altering the architecture of the outer Solar System by changing the configuration of the giant planet masses gives us a more detailed look at the specific role played by the planets in molding the OC. As well as providing clues to the dynamical processes of its formation, the structure of the OC will determine its response to external perturbations, particularly the effect on the flux of objects into the inner Solar System.

\begin{figure*}[!ht]
  \begin{center}
    \epsscale{0.8}
    \plotone{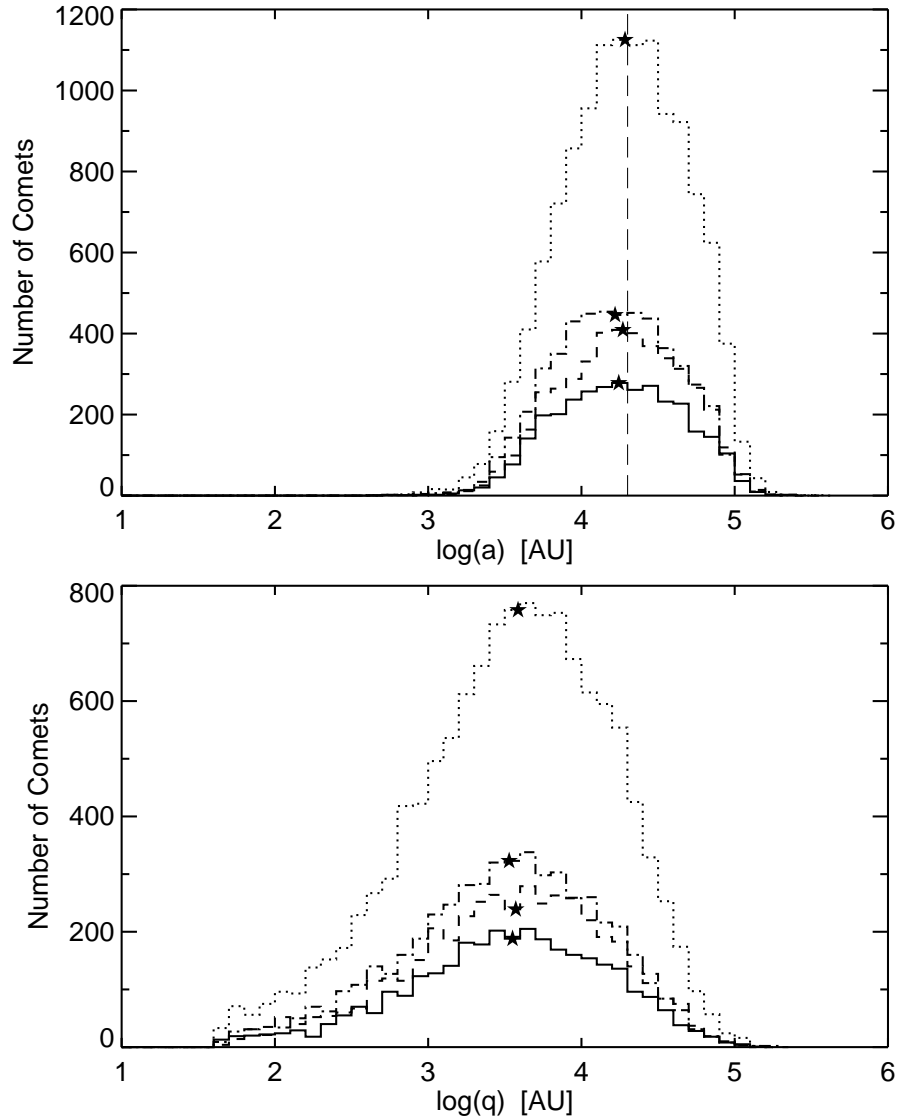}
  \end{center}
  \caption{The distribution of OC objects in semimajor axis ($a$) and perihelion ($q$) at the end of the simulation. Stars indicate the median values of each distribution. The vertical dashed line delimits the boundary between the inner and outer OC at 20,000 AU. Solid: \control, Dashed: \twoSaturns, Dash-dot: \iceSaturn, Dotted: \allIce}
  \label{fig:OCadist_end}
\end{figure*}

To examine the overall structure, we look at the distributions in semimajor axis and perihelion distance for each OC after 4.5 Gyrs, plotted in Figure~\ref{fig:OCadist_end}. For reference, we have over-plotted a line at 20,000 AU on the semimajor axis distribution to show the location of the boundary between the inner and outer OC. It is evident that the distributions are all very similar: most particles are found near the inner-outer boundary and the number declines with distance from that boundary. The distributions all have long tails at small distances, which is likely due to close stellar passages that strongly alter the orbits of some particles. The median $a$ differs by no more than 15\% between the four configurations. What is perhaps most surprising, though, is that the \iceSaturn\ run has the smallest median semimajor axis at $a=16,520$ AU. This may suggest that a smaller mass Saturn is not only more efficient at sending particles to the OC, but it also preferentially places those particles on tighter orbits. A reduced mass Jupiter places objects on more extended orbits, which can be seen in the \twoSaturns\ and \allIce\ runs where the median location is $a=18,450$ AU and $a=19,230$ AU, respectively. The \control\ run median value occurs at $a=17,460$ AU.

The perihelion distribution shows a similar trend. The peak of each distribution occurs in approximately the same place, at about 4000 AU. The range in median $q$ is small, only a 12\% variation between simulations. Once again, the \iceSaturn\ configuration creates an OC in which the particles have the smallest perihelion distance. The median occurs at $q = 3360$ AU. The largest value is found in the \allIce\ run where the median is $q=3840$ AU. This distribution has a more significant tail toward shorter distances, which is reflected in the fact that the peak of the distribution occurs at a larger distance than the median perihelion values. While the same is true of the semimajor axis distribution, the spread is somewhat offset by the larger distances. The \control\ and \twoSaturns\ runs have median $q=3560$ AU and 3720 AU, respectively.

\begin{figure*}[!t]
  \begin{center}
    \epsscale{1.11}
    \plottwo{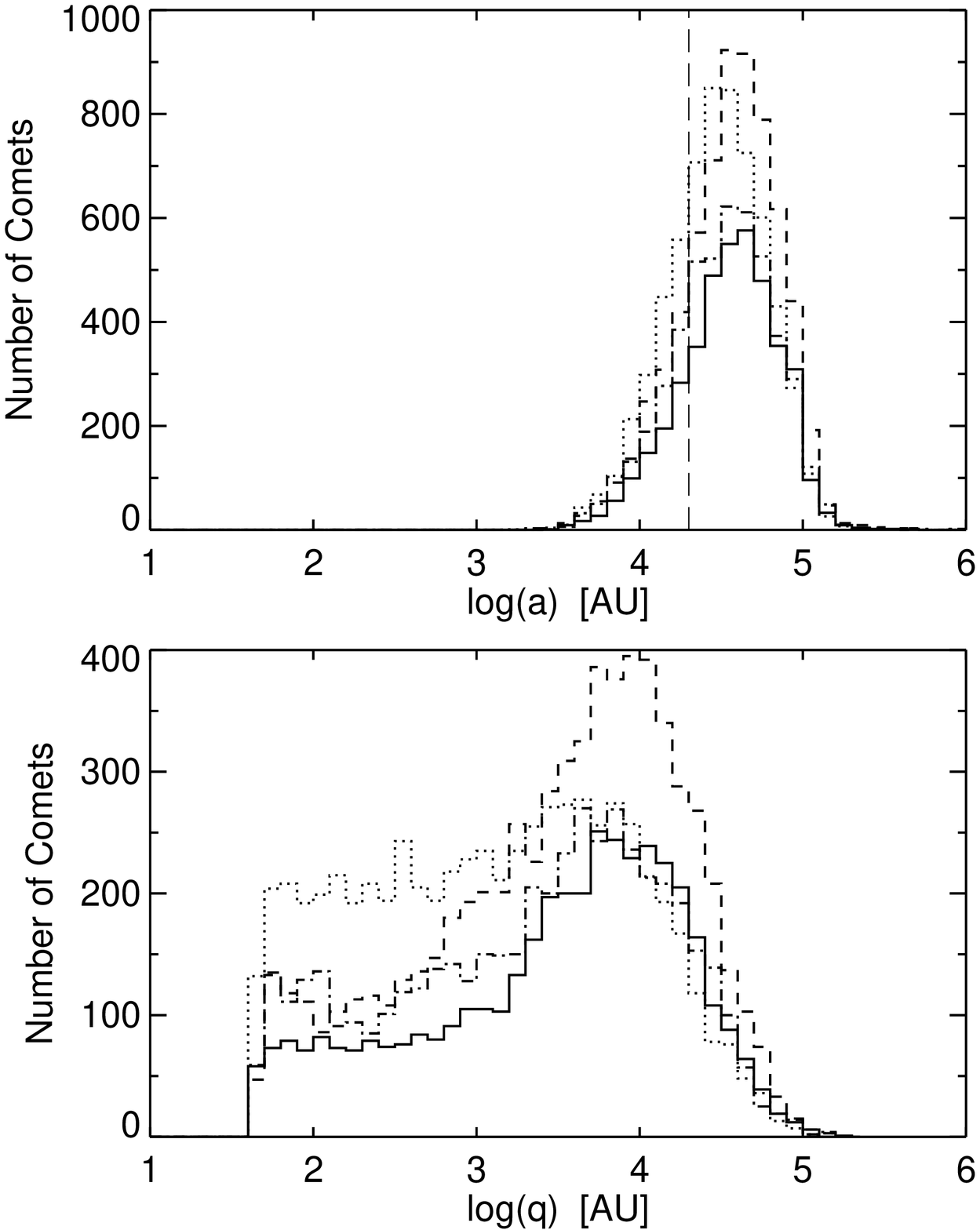}{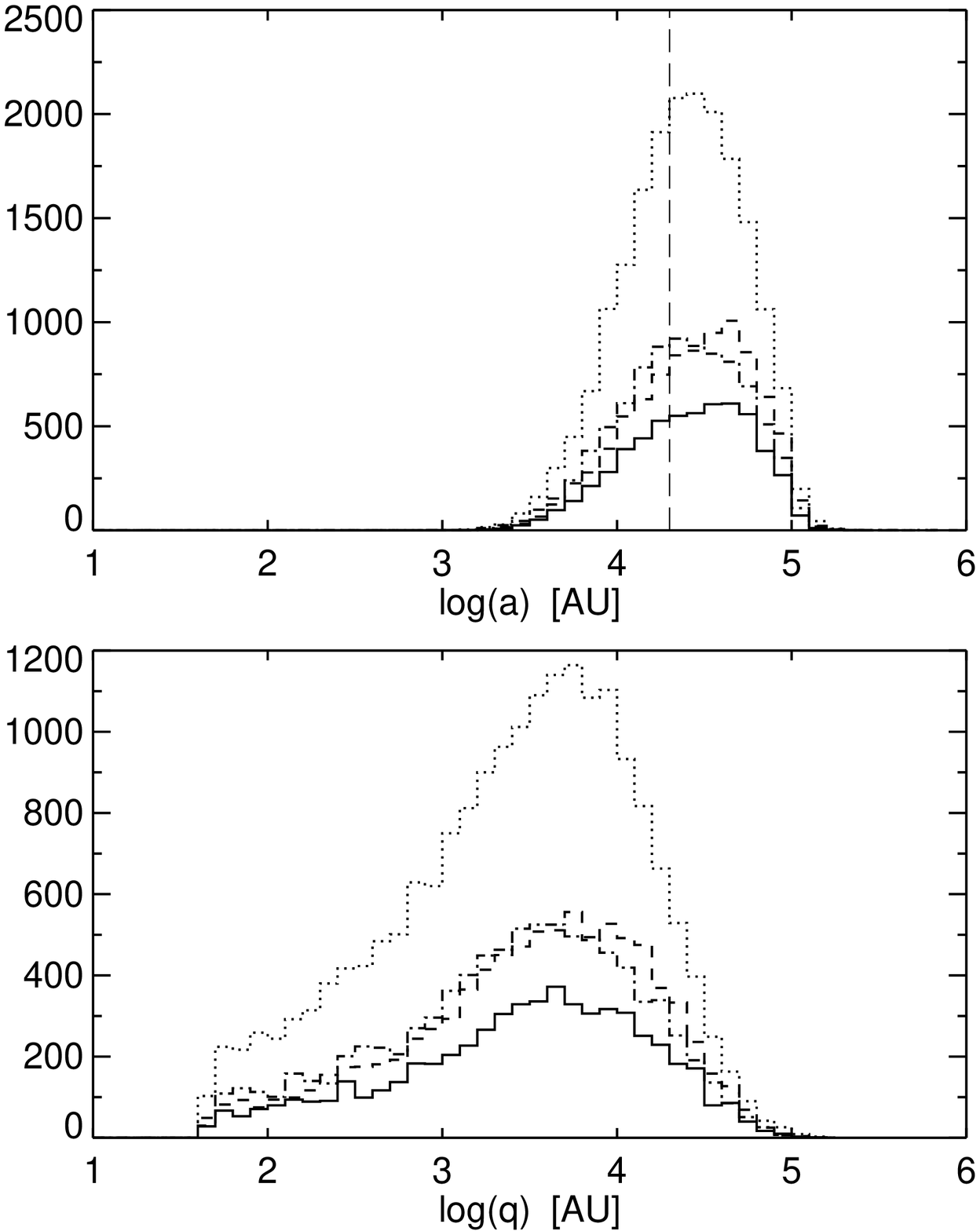}
  \end{center}
  \caption{Same as Figure~\ref{fig:OCadist_end} but at 200 Myrs (left) and 1 Gyr (right). The median value markers have been left off for clarity.}
  \label{fig:OCadist_mid}
\end{figure*}

The evolution of the overall structure is significant and further underscores the importance of the 4.5 Gyr integration time. In Figure~\ref{fig:OCadist_mid}, we plot the semimajor axis and perihelion distributions at 200 Myrs and 1 Gyr. The structural evolution of each OC expands on the evolution seen in the TE (Figure~\ref{fig:OCtrapEff}) and the inner-outer OC ratio (Figure~\ref{fig:OCratio}). When particles are first sent to the OC, they primarily occupy more extended orbits in the outer OC. With time, the median value moves inward and the number of particles increases to its peak and then decreases. In the semimajor axis distribution, we see that although the location and number of particles in each OC changes with time, the overall structure is very similar between the four simulations and they evolve in the same way. The perihelion distribution, on the other hand, does not evolve in such a uniform way. A buildup of particles at small $q$ occurs at earlier times as a result of the path by which particles become OC objects, in which the semimajor axis must diffuse outward before particle perihelion can be altered. This process will be discussed in Section~\ref{subsec:timescales}.

We therefore conclude that the masses of the giant planets play a significant role in determining the size of the OC, that is the number of comets and hence the mass, but have little effect on its final structure. The differences in number certainly distinguish the distributions one from another; however, qualitatively they are very similar. Instead, the structure is primarily influenced by the Galactic environment \citep[e.g.,][]{Brasser2007, Kaib2008, Kaib2009, Brasser2010, Levison2010}.

\subsection{Flux of Observable LPCs}
\label{subsec:flux}

Once particles have made it into the OC, they may be affected by external forces which perturb them onto Earth-crossing orbits. As we have shown, the number of particles that are transferred onto such orbits is influenced by the total number of particles and their specific locations in the OC. These particles are of particular interest due to their potential for impacts. We examine the extent to which the giant planets protect the terrestrial planets from impact by studying the flux of particles into the inner Solar System (inside 2AU) at 250 Myrs and 500 Myrs from the present. We use the cloned simulations to evaluate these numbers.

\begin{deluxetable*}{lccccccc}
    \tablecolumns{8}
    \tabletypesize{\footnotesize}
    \tablewidth{0pt}
    \tablecaption{Flux of Particles Inside 2 AU}
    \tablehead{
        \colhead{} & 
        \multicolumn{3}{c}{4.5--4.75 Gyrs} & 
        \colhead{} & 
        \multicolumn{3}{c}{4.75--5.0 Gyrs} \\
        \cline{2-4} \cline{6-8} \\
        \colhead{Simulation} & 
        \colhead{Total} & 
        \colhead{Percent} & 
        \colhead{Fraction of OC} &
        \colhead{} &
        \colhead{Total} & 
        \colhead{Percent} & 
        \colhead{Fraction of OC} \\ 
        \colhead{} & 
        \colhead{Number} & 
        \colhead{of OC} & 
        \colhead{per year($\times10^{-12}$)} &
        \colhead{} &
        \colhead{Number} & 
        \colhead{of OC} & 
        \colhead{per year ($\times10^{-12}$)}
  }
  \startdata
      control   & 165  & 0.089 & $3.56$ & & 262  & 0.146 & $5.84$ \\
      2Saturns  & 245  & 0.132 & $5.29$ & & 264  & 0.147 & $5.89$ \\
      iceSaturn & 323  & 0.174 & $6.95$ & & 292  & 0.162 & $6.47$ \\
      allIce    & 811  & 0.430 & $17.2$ & & 645  & 0.352 & $14.1$
  \enddata
  \label{tab:flux}
\end{deluxetable*}

We define the flux as the total number of objects that come inside 2AU over a 250 Myr period. We have examined the flux of particles in each system at two separate times to see how the flux changes with time. Our first window is from 4.5--4.75 Gyrs and the second is from 4.75--5.0 Gyrs. For each time frame, we count the total number of objects injected inside 2 AU, calculate these numbers as a percentage of the OC at the end of the 250 Myr window, and convert that to a fraction of the OC injected inside 2 AU per year. These numbers are given in Table \ref{tab:flux}. When we change planetary mass configurations from the control, we expect the total flux to increase because planets with smaller mass have less influence on passing particles. During both windows, the similarity in numbers for the first three simulations indicates that the Saturn mass is actually the interesting cutoff mass.

When comparing simulations, the relative flux remains fairly steady--the \control, \twoSaturns, and \iceSaturn\ flux numbers are all within a factor of two while the \allIce\ configuration has 3--5 times more particles at both times. The similarity of the percentages of the first three simulations over time, and the increase in the \allIce\ simulation, once again emphasizes that Saturn plays as significant a role as Jupiter in this process. Flux increases as planetary mass decreases, demonstrating that the outer giant planets do act as ``planetary protectors'', shielding the inner Solar System from OC objects.

\begin{figure*}[!th]
  \begin{center}
    \epsscale{0.9}
    \plotone{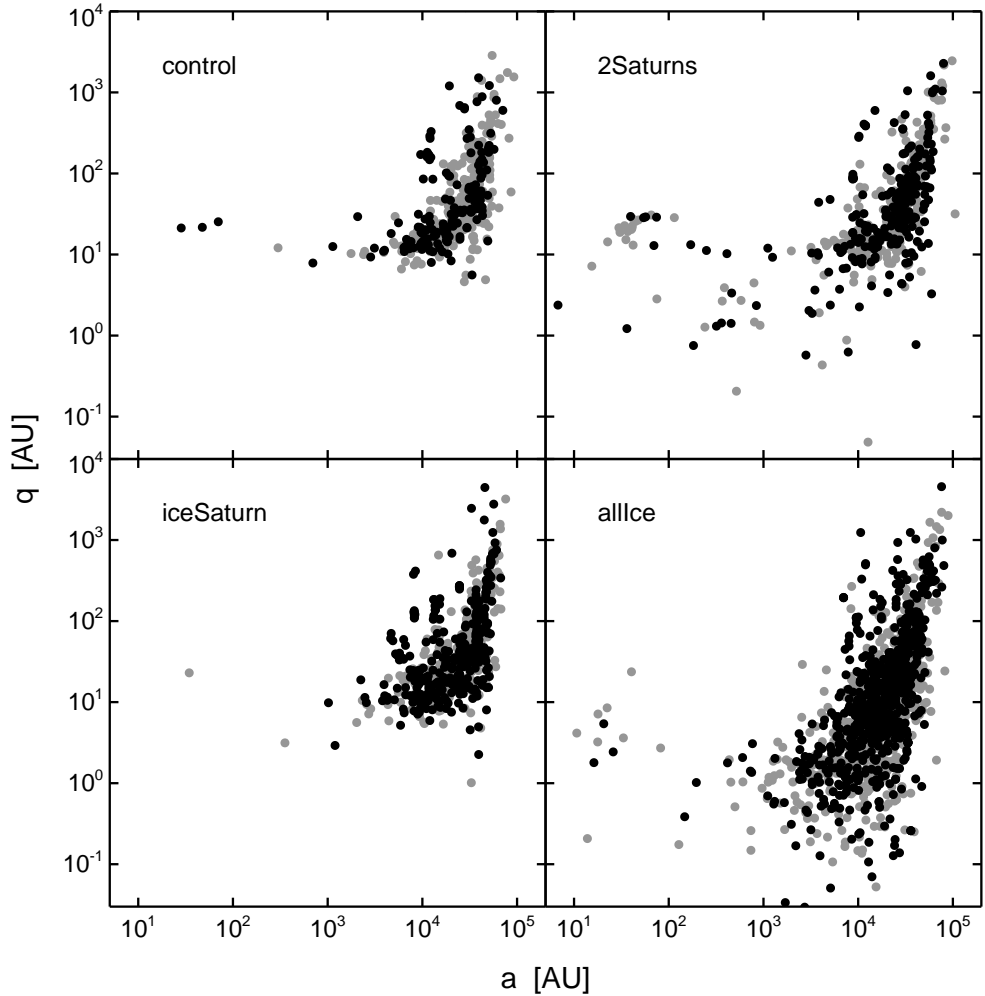}
  \end{center}
  \caption{Perihelion distance as a function of semimajor axis for each particle that comes inside 2 AU in the cloned simulations. Particles that enter the inner Solar System from 4.5--4.75 Gyrs are shown as gray circles while those from 4.75--5.0 Gyrs are shown in black. Each point represents an individual particle 10 Myrs before it enters the inner Solar System. This occurs at a different time for each particle.}
  \label{fig:flux}
\end{figure*}

To further examine these particles, we follow their evolution from their initial location in the protoplanetary disk to the point at which they entered the inner Solar System, ending either with their ejection by one of the planets or the end of the simulation. In Figure~\ref{fig:flux}, we plot perihelion distance as a function of semimajor axis for each particle. These locations are found 10 Myrs before the particle reaches 2 AU on its journey into the inner Solar System, which occurs at a different time for each particle. We choose to show the locations at 10 Myrs out because not only is the orbital period of objects in the outer OC of order 10 Myrs, but these particles also show significant evolution over the past 10 Myrs as opposed to that between 10 and 20 Myrs. Within 5 Myrs, there is short term orbital evolution as the particles make their way into the inner Solar System. The plot shows that as the total mass of the inner two planets decreases, the particles that are on their way into the inner Solar System come from a distribution of orbits that extends to smaller semimajor axis and perihelion. This further highlights the distinction between systems that have Saturn-mass planets and those that lack them. Interestingly, in the \twoSaturns\ and \allIce\ runs, we appear to have trapped a couple of Halley-type comets; however, the numbers are too small for us to say anything about the production rate of these comets as a function of outer Solar System architecture.

We find that these particles show no preference for initial protoplanetary disk location; they are uniformly spread across the range from 4--40 AU. This is true at 4750 Myrs and 5000 Myrs and suggests that altering the planetary mass configuration does not affect which particles end up in the OC. We also find that the inner OC contributes substantially to the flux of comets into the inner Solar System, confirming the result from \citet{Kaib2009}.

\section{Discussion}
\label{sec:discussion}

\subsection{Timescales}
\label{subsec:timescales}
We have examined properties of the OC produced in each of our simulations to quantify the role played by the outer planets in creating and altering each comet cloud. A full examination of the formation and structure of the OC, however, must include an analysis of the events that influence the path of a particle as it evolves onto an OC orbit. We shall discuss here the period, energy diffusion time, and tidal torquing time of the particles as they interact with the planets, stars, and the Galactic tide. These timescales are important because they help determine the location of boundaries in the OC as well as define the different stages of the orbital evolution of a comet. 

\begin{figure}[!tb]
  \begin{center}
    \epsscale{1.0}
    \plotone{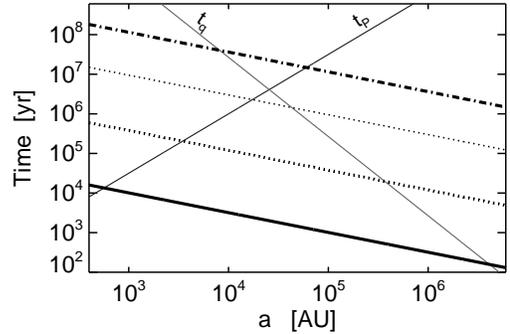}
  \end{center}
  \caption{Timescales for the scattering and torquing of test particles as a function of semimajor axis. Unmarked lines are energy diffusion times due to the planets, calculated as in Equation \ref{eq:t_d}. We have included four example lines that span the range covered by the combination of masses and semimajor axes of the planets in the simulations. Solid thick line: \control\ Jupiter, thick dotted line: \control\ Saturn, thick dash-dot line: \control\ Neptune, thin dotted line: \iceSaturn\ Saturn}
  \label{fig:timescales}
\end{figure}

We first look at these timescales to understand the boundary between the inner and outer OC. Traditionally, this boundary has been set by the intersection between the period and the tidal torquing time---the particle must attain a large enough distance such that its orbit can be externally perturbed by passing stars or the Galactic tide but small enough so that it still remains bound to the Sun. The orbital period of a comet is given by $t_P=(a $ AU$)^{3/2}$ yr. For the tidal torquing time, $t_q$, we follow \cite{DQT1987}:
\begin{equation}
  t_q = 2.67 \times 10^7 \left(\frac{\Delta q}{10 \: \mathrm{AU}}\right) \left(\frac{25 \: \mathrm{AU}}{q}\right)^{1/2} \left(\frac{10^4 \: \mathrm{AU}}{a}\right)^2 \: \mathrm{yr}
  \label{eq:t_q}
\end{equation}
where $\Delta q$ is the typical perihelion change, $q$ is the perihelion distance, and $a$ is the semimajor axis. The equation is defined using $\Delta q = 10$ AU because a change of this magnitude implies that the diffusion timescale of the particle (Equation \ref{eq:t_d}) is set either by a different planet or by no planet at all if it has moved beyond the planetary region; as a result, the particle orbit will evolve in a much different way. We have plotted $t_q$ in Figure~\ref{fig:timescales} for $\Delta q = 10$ AU and $q = 25$ AU. As can be seen in Figure~\ref{fig:timescales}, $t_q = t_P$ at \about$20,000$ AU. While \cite{Oort1950} originally hypothesized that this distance was the inner edge of the comet cloud, \cite{Hills1981} showed that this was simply a selection effect due to the rarity of stellar passages in this region. As a result, this distance is now quoted as the boundary between the inner and outer OC \citep{Hills1981,DQT1987,Dones2004,Kaib2008,Kaib2011b}.

The diffusion time, $t_d$, is the timescale on which a planet changes the semimajor axis of a comet by a factor of two. We again follow \cite{DQT1987} though we adopt the formulation in \cite{Brasser2010}:
\begin{equation}
  t_d = \frac{2.98 \times 10^4}{\beta^2} \left(\frac{10^4 \: \mathrm{AU}}{a}\right)^{1/2} \bigg(\frac{a_p}{5 \: \mathrm{ AU}}\bigg)^2 \left(\frac{m_p}{\Mearth}\right)^{-2} \: \mathrm{ Myr}
  \label{eq:t_d}
\end{equation}
where $m_p$ and $a_p$ are the mass and semimajor axis of the scattering planet and $\beta$ is a measure of the energy change experienced by a comet during each passage, defined by $\Delta(1/a)=\beta \left(m_p/M_{\odot}\right) \left(1 \: \mathrm{AU}/a_p\right)$. We use the traditional value of $\beta=10$ \citep{Fernandez1981, DQT1987, Tremaine1993}.

The intersection of these timescales can be used to understand the perturbation mechanisms that affect the orbit of a comet at various stages in its evolution. Comets initially diffuse outwards in $a$ as a result of interactions with the planets while their $q$ remains in the planetary region. At the semimajor axis where the energy diffusion time due to perturbations from a given planet becomes equal to the tidal torquing time ($t_q = t_d$), the comet will be lifted out of the planetary region to larger $q$ and into the OC \citep{DQT1987}. In Figure~\ref{fig:timescales}, this is represented as follows: comets follow lines of $t_d$ at small $q$. When $t_d$ crosses $t_q$, \ie the energy diffusion time exceeds the tidal torquing time, the comet is lifted out of the planetary region and into a safely bound orbit. However, comets may be ejected from the system before they reach the OC if they receive large enough energy kicks from the planets. If the line $t_d$ crosses $t_P$ before it crosses $t_q$, then that comet is much more likely to be ejected from the system on its next passage through the planetary region. As a result of this path, \cite{DQT1987} concluded that this location should set the inner edge of the comet cloud. 

By this logic, Jupiter and Saturn have no impact on the comet cloud inner edge. In order for a planet to influence the location of the inner edge, the semimajor axis at which the energy diffusion timescale of the planet equals the tidal torquing timescale must be less than the semimajor axis where the diffusion timescale matches a comet's orbital period; that is, $a_{t_d=t_q} < a_{t_d=t_P}$. If this is true, then the planet can send a comet to the OC. In Figure~\ref{fig:timescales}, this occurs for lines of $t_d$ that cross $t_q$ before $t_P$. In our simulations, this is true only of Uranus and Neptune. While we did not alter the mass of Neptune in any of the simulations, we can see from Equation~\ref{eq:t_d} that reducing the mass increases the timescale at a given $a$, so the line moves up on the plot as mass decreases. We have also not accounted for planetary migration in these simulations---Uranus and Neptune likely accumulated their mass closer to the Sun and then migrated outward \citep{Thommes1999}. If we decrease $a_p$ in Equation~\ref{eq:t_d}, the timescale increases; however, the semimajor axes of Uranus and Neptune likely only changed by a factor of 2--3 \citep{Tsiganis2005}, which would change $t_d$ by a factor of 4--9 and would not have a sizable effect on the relative significance of the planets. We can therefore conclude that Jupiter and Saturn have no bearing on the location of the inner edge of the OC. 

The inner boundary of the OC is not well constrained. Previous studies have placed it anywhere from 100 AU to 10,000 AU with an average of a few thousand AU \citep[e.g.,][]{DQT1987, Dones2004, Kaib2009, Brasser2012}. It has also been noted that the density of the Solar environment as well as the stochastic effects of stellar perturbations can contribute to this large range \citep{Brasser2010,Kaib2011b}. If we use the simple argument that the location of the inner edge of the OC is the smallest semimajor axis where the energy diffusion time for a given planet is equal to the tidal torquing time, then Neptune sets the inner edge and in our simulations it occurs at $a \simeq 8000$ AU. Given that we have a fairly large number of objects at smaller semimajor axes in each of our simulations, it is possible that due to the stochasticity of stellar perturbations, single close encounters could push orbits inward. It is important to note that while we are able to assert that Jupiter and Saturn do not affect the inner edge, our simulations do not allow us to definitively say that Neptune does set the location. Further experiments in which Neptune's mass is altered are needed to make that assertion. We plan to do this in future work.

\subsection{Planetary Roles}
\label{subsec:planet_roles}

Our primary aim has been to determine the extent to which the outer planets helped to shape the OC as well as their ability to protect the inner planets from incoming OC comets. We have seen that the mass configuration of the giant planets does not drastically alter the structure of the OC (Figures~\ref{fig:OCratio}~and~\ref{fig:OCadist_end}). Systems with less massive planets create an OC with more comets, but the overall distributions of comets in $a$ and $q$ are very similar. 

\citet{Hills1981} suggested that the inner OC is a source for comet showers, periods of very high cometary flux on Earth-crossing orbits. If a large number of comets live in the inner OC, then close stellar encounters that trigger comet showers will push proportionally more comets into the inner Solar System, resulting in more intense showers. On the other hand, if more of the comets live in the outer OC, showers would be less intense; however, stellar encounters will more efficiently unbind these comets from the Sun, decreasing the overall TE. Figure~\ref{fig:OCratio} shows that the relative numbers in the inner and outer OC does not change dramatically as a function of planetary mass. The total number in each cloud increases as planetary mass decreases, which would suggest more intense comet showers, simply because there are more comets in the cloud to be perturbed. We can conclude that the similarity in structure between each OC is likely a result of perturbation mechanisms; the pathways for entry into and exit from the cloud are the same in each case.

Despite these overall similarities, we can draw some distinctions between the planets, picking out their importance in the overall evolutionary process.

\subsubsection{Jupiter vs. Saturn}
\label{subsubsec:jupiter_saturn}
Particles in our simulations experience gravitational interactions with the planets. The more massive the planet, the greater the gravitational force exerted on passing particles. Because Uranus and Neptune are least massive, particles can more easily pass through that region without experiencing major perturbations. Particles that move into the Jupiter-Saturn region are more likely to interact with these planets, which generally leads to ejection from the system. It is because of this that earlier studies considered Jupiter to be important for shielding purposes \citep{Safronov1972, Wetherill1994, Horner2010}. However, we see here that Saturn plays an equally significant role. We can compare the relative importance of Jupiter and Saturn by examining the \twoSaturns\ and \iceSaturn\ configurations in which Saturn and Jupiter, respectively, have the dominant roles.

In the two simulations, the TEs into the OC are comparable. This is not surprising because most OC particles are placed on those orbits by Uranus or Neptune, not by Jupiter or Saturn (see Table~\ref{tab:pl_percents}). If the Jupiter mass were the important mass, then the results of the \control\ and \iceSaturn\ runs should be similar. Instead, we see that the TE is slightly higher when there is no Saturn. The natural thought would be that this is because a reduced-mass Saturn can aid in placing particles into the OC rather than ejecting them entirely from the system because the boost in energy it provides is not too large; however, in the \iceSaturn\ system, the smaller Saturn actually sends fewer particles into the OC. This is also true of the Jupiter-mass planet in the \iceSaturn\ run. When we reduce the mass of these two giant planets, fewer particles are scattered to lower perihelia and interact with them, whether they be sent to the OC or ejected from the system entirely. Instead, more of the particles are affected by Neptune and the size of the OC increases.

We find that the number of particles that enter the inner Solar System is approximately the same in the \control, \twoSaturns, and \iceSaturn\ runs. The \iceSaturn\ comet cloud TE is higher, though, so it has a larger reservoir from which particles may be perturbed. If instead we look at the number of particles that come inside 2AU as a percentage of the number total number of OC particles at the end of the simulation, the numbers are very similar and become even more so as the system evolves. Because we see a much more significant increase in both the number and percentage of particles in the \allIce\ run, we conclude that the protection provided by the outer giant planets is unchanged until the system no longer contains a Saturn-mass or larger planet. 

K-S tests of the semimajor axis distributions show that the interesting cases are the \control-\allIce, the \twoSaturns-\iceSaturn, and the \iceSaturn-\allIce\ comparisons. Despite the fact that the distributions look qualitatively similar in overall structure, they are different enough to be quantitatively distinguishable. The significance of the K-S statistic is of order $10^{-5}$, $10^{-8}$, and $10^{-15}$, respectively; in all these cases, it is very unlikely that the two populations were drawn from the same parent population. While having a Jupiter-mass planet makes a difference, it is not as significant as when there is no massive planet at all. Whether a system has a Saturn-mass planet, a Jupiter-mass planet, or both, the evolution of the system, the number of Earth-crossing particles, and the resulting structure of the OC is similar. When there is no longer at least a Saturn-mass planet, the differences are much more drastic.

\subsubsection{Neptune}
\label{subsubsec:neptune}
We argued in Section~\ref{subsec:timescales} that Neptune plays an important role because it sets the location of the inner edge of the OC in our simulations. We can underscore the significance of Neptune by examining the planet of last encounter for each OC particle. If we assume that the last planet encountered by each particle is the one that kicked it into the OC, then in each simulation the majority of the particles are indeed kicked out by Neptune. Table~\ref{tab:pl_percents} lists the fraction of OC particles at the end of each simulation that last encountered each planet. As the planet that is the furthest out and almost the least massive, Neptune plays a major role in shaping the OC because it is responsible for much of the population. As a particle's perihelion evolves inward, it is likely to encounter Neptune first; if it does so, it is more likely to end up in the OC than if it had encountered Jupiter or Saturn. In systems where planetary mass does not decrease with distance from the central star, the result could be very different.

\begin{deluxetable}{lcccc}
  \tablecolumns{5}
  \tablewidth{0pt}
  \tabletypesize{\footnotesize}
  \tablecaption{Planet of Last Encounter at 4.5 Gyr}
  \tablehead{
    \colhead{Simulation} & 
    \colhead{Jupiter}    & 
    \colhead{Saturn}     & 
    \colhead{Uranus}     & 
    \colhead{Neptune}    
  }
  \startdata
  control   & 0.0200  & 0.0877 & 0.1221 & 0.7432\\
  2Saturns  & 0.0252  & 0.1317 & 0.1551 & 0.6835\\
  iceSaturn & 0.0125  & 0.0676 & 0.1249 & 0.7886\\
  allIce    & 0.0455  & 0.0973 & 0.2053 & 0.6489
  \enddata
  \label{tab:pl_percents}
\end{deluxetable}

\subsection{Implications for the Protoplanetary Disk}
\label{subsec:protodisk}

Understanding the role of the planets in the evolution of the Solar System not only elucidates the properties of the OC but may also shed some light on an inconsistency seen in many OC models. The number and mass of objects in the OC is related to the mass of the protoplanetary disk. In simulations, we can directly connect the two masses because we know the TE of each OC. Using the results from our \control\ run and an estimate for the flux of observable LPCs as 1--3 per year \citep{Oort1950, Everhart1967,Heisler1990}, we compute an OC with $3-8\times10^{11}$ objects, which is comparable to other estimates in the literature of  $10^{11}-10^{12}$ objects \citep{Oort1950, Heisler1990, Kaib2009}. In order to account for this number of objects, a low TE implies a more massive protoplanetary disk while a larger TE suggests a less massive disk. Results from our \control\ run and previous studies produce TEs (see Section \ref{subsec:trapeff}) that require the disk mass to be larger than is believed plausible.

\citet{Kaib2009} suggested that the majority of observed LPCs may originate in the inner OC. This could help resolve the mass discrepancy because the inner OC is easier to populate than the outer OC.  If many comets originate in the inner OC and it is more massive than the outer OC, the required disk mass could be dramatically decreased. The planets most efficiently populate the inner OC when they are less massive; therefore, in the beginning stages of the formation of the Solar System, more objects could be sent into the inner OC than have been previously acknowledged. Although most gas giants form within 10 Myrs \citep{Dangelo2011}, during this period they spend most of their time as ice giant-sized objects. The smaller mass of the planets during this short period of time could elevate deposition of particles into the OC and reduce the calculated discrepancy in the mass of the initial protoplanetary disk.

\subsection{Future Work}
\label{subsec:future}
The formation and evolution of the OC is tied up in the larger history of the Solar System, a complex story that itself is still being pieced together. There are many factors to consider, from the original birth environment of the Sun and the possibility of stellar migration, to the formation of the giant planets and planetary migration through the protoplanetary disk, to accretion of the terrestrial planets. 

There are clearly limitations to this work. We have chosen to leave the planets at a single mass and on their current orbits for the duration of the simulation. This greatly simplifies the calculation while allowing us to begin to see how differences in planetary mass affect the end result. This work would also be remarkably improved if set in the context of a more realistic Solar System evolution model, such as the Nice model \citep{Tsiganis2005} and the Grand Tack model \citep{Walsh2011}, which offer compelling pictures of the early evolution of the Solar System. This, however, represents a significant leap from the current model. A reasonable first step would be to implement planetary migration in the existing OC evolution model. This is especially important because the period of planet growth is very short compared to the age of the Solar System \citep[e.g.,][]{Lissauer1987}, so the influence of planetary mass, particularly decreased mass, plays out in a very small window. Migration may also lead to movement of planets through resonances, which has been shown to have distinct and drastic effects on the nearby small body populations \citep{Gomes2005_LHB}. Such movement could improve efficiency of OC population. Taken alone, these simulations do not create an excess of OC objects within the first 10--100 million years when the outer planets are growing to their current size. If is likely that if placed in a denser solar environment or if planetary migration---which the timing of the late heavy bombardment suggests may last up to 700 Myrs---is included in the models, the results would be promising. 

Barring such changes, the current model can be used to continue examination of the role of the outer planets. The delivery of volatiles into the inner Solar System is affected by the properties of the outer planets \citep{Raymond2006}. Studying the impact of giant planet mass on the injection of volatiles to inner terrestrial planets would be illuminating not only for studies of our own Solar System, but could have important ramifications for extrasolar planetary systems as well.

\section{Conclusions}
\label{sec:conclusions}

We have examined how the architecture of the outer Solar System helped to mold the structure and formation of the OC. To do so, we ran four simulations with different planetary mass configurations. The simulations include 100,000 massless particles that interact with the four giant planets, the Sun, passing stars, and the Galactic tide. This study differs from previous studies that have examined the role of the planets in that we allow the systems to evolve over the full 4.5 Gyr history of the Solar System. We have shown that integration over this full period of time is necessary to achieve a more accurate understanding of planetary roles. Not only does the TE of each OC change with time, but the overall structure changes fairly dramatically. Early in the evolution, particles are deposited into the outer OC; over time, more particles are found in the inner OC.

Changing the mass of the four outer planets does not significantly alter the overall structure of the OC but does impact the number of particles that end up as OC objects at the end of the simulation. Each OC shows the same qualitative distribution in semimajor axis, but the \allIce\ run has significantly more particles in the OC than the other three simulations. As the amount of mass in the system decreases, the overall TE increases; the \control\ run, which mimics the Solar System as it is today, has the lowest trapping efficiency, while the \allIce\ run, in which each of the planets has the mass of Neptune, has a trapping efficiency that is five times higher. TEs in the \twoSaturns\ and \iceSaturn\ runs are very similar, showing that the contributions from Jupiter and Saturn are of approximately equal magnitude. Although the effect may be minimal, taken together these results indicate that accounting for planet formation in the early stages of the evolution of the system may help to increase the overall TE and contribute to a solution to the problem that a low TE requires a large protoplanetary disk.

Additionally, the existence of the gas giant planets does reduce the flux of comets into the inner Solar System, specifically reducing the number of particles that come within 2 AU of the Sun on Earth-crossing orbits. Similar to the results of \citet{Horner2010}, we find that the flux difference between configurations is less than an order of magnitude. This is much smaller than the factor of 1000 calculated by \citet{Wetherill1994}, though we emphasize that based on simplifications in the calculation, his numbers more accurately describe short period comets rather than LPCs. Once again, Jupiter and Saturn provide equal protection, in that when each planet's mass is independently reduced, the number of particles that come inside 2 AU is approximately the same. This implies that while having a Jupiter-mass planet in the system is certainly helpful, a Saturn-mass object provides sufficient protection for the inner planetary region.

Finally, although our simulations indicate that Neptune plays a critical role in determining OC structure, additional simulations varying Neptune's mass must be run for confirmation. This will provide further insight into the role of the outermost planet in any system. Data from these simulations as well as inclusion of new models that more accurately reproduce the early stages of Solar System formation, including planetary growth and migration, will promote enhanced study of the OC and lead to further constraints on its formation and evolution.

\acknowledgements
We would like to thank the anonymous referee for helpful comments that improved the clarity of this paper. Support for this research was provided by the NSF through grant AST-0709191.

\end{document}